\documentclass[aps,showpacs,prb,superscriptaddress,floatfix,twocolumn,citeautoscript]{revtex4-1}
\usepackage{graphicx}
\usepackage{amsmath}
\usepackage{amssymb}
\usepackage{float}
\usepackage{bm}
\usepackage{dcolumn}
\usepackage{subfigure} 
\usepackage{hyperref}
\usepackage[usenames]{color}
\usepackage{multirow}
\usepackage{lipsum}
\usepackage{afterpage}
\usepackage{subfigure}
\usepackage{program}
\hypersetup{pdfborder=0 0 0,colorlinks=true,citecolor=blue,linkcolor=blue}
\input epsf
\DeclareGraphicsExtensions{.pdf,.jpeg,.png, .eps}

\usepackage{graphicx}
\usepackage{amsmath}
\usepackage{amssymb}
\usepackage{float}
\usepackage{bm}
\usepackage{dcolumn}
\usepackage{subfigure} 
\usepackage{hyperref}
\usepackage[usenames]{color}
\usepackage{multirow}
\usepackage{subfigure}
\hypersetup{pdfborder=0 0 0,colorlinks=true,citecolor=blue,linkcolor=blue}
\input epsf
\DeclareGraphicsExtensions{.pdf,.jpeg,.png, .eps}

\begin{document}
\title{Tuning electronic properties and contact type in van der Waals heterostructures of bilayer SnS and graphene   }
\author{M.R. Ebrahimi}
\affiliation{{Department of Physics, Shahid Beheshti University, Evin, Tehran 1983969411, Iran}}
\author{T. Vazifehshenas}
\email{t-vazifeh@sbu.ac.ir}
\affiliation{{Department of Physics, Shahid Beheshti University, Evin, Tehran 1983969411, Iran}}

	\begin{abstract}
Using first-principles calculations, we study the structural and electronic properties of the bilayer SnS/graphene, bilayer SnS/bilayer graphene (AA-stacked), bilayer SnS/bilayer graphene (AB-stacked) and monolayer SnS/graphene/monolayer SnS van der Waals (vdW) heterostructures. Electronic properties of all components of the vdW heterostructures are well preserved, which reflects the weakness of the vdW interaction. In the cases of  bilayer SnS/graphene and bilayer SnS/bilayer graphene (AA-stacked), an Ohmic contact is formed which can be turned first into p-type and then into n-type Schottky contacts via application of an external electric field. Calculations show that an Ohmic contact is also formed at the interface of bilayer SnS/bilayer graphene (AB-stacked) heterostructure, but interestingly, by applying the perpendicular electric field a transition from semimetal/semiconductor contact to semiconductor/semiconductor one occurs which can enhance its optical properties. Alternatively, in the monolayer SnS/graphene/monolayer SnS vdW heterosructure, a p-type Schottky contact is established that changes into Ohmic contact under an applied electric field. Our results clearly indicate that the electronic properties of the vdW heterostructures can be tuned efficiently by external electric field, which is important in designing of new nanoelectronic devices.

	\end{abstract}
	\date{\today}	
    \maketitle{\textbf{\textit{Keywords : }}}
	
\section{Introduction}

Since the discovery of atomically thin two-dimensional (2D) graphene exhibiting intriguing physical properties \cite{K. S. Novoselov,A. K. Geim}, 2D materials have received extensive attention from the researchers in the fields of condensed matter physics and material science. Graphene is a semimetal with ultrahigh carrier mobility  \cite{X. Du}, but the lack of a band gap has limited its application in nanoelectronic devices. In the recent two decades, a wide variety of 2D materials such as hexagonal boron nitride (h-BN) \cite{Y. Shi,L. Song,K. S. Novoselov_2}, silicene \cite{S. Cahangirov,C.-C. Liu,A. Kara,P. Vogt}, graphitic carbon nitrides \cite{X. Wang,Q. Hu,S. Cao}, transition metal dichalcogenides (TMDCs) \cite{J. N. Coleman,B. Radisavljevic,Q. H. Wang,M. Chhowalla} and phosphorene \cite{H. Liu,L. Li,S. P. Koenig} have been intensively investigated and successfully fabricated.

In particular, phosphorene a single layer of black phosphorus (BP) with an orthorhombic crystal structure exhibits fascinating anisotropic optoelectronic \cite{V. Tran,F. Xia,S. Saberi-Pouya} and transport \cite{J. Qiao} properties. The carrier mobility of phosphorene is as high as 1000 cm$^{2}$V$^{-1}$S$^{-1}$ \cite{M. Akhtar}, which is much larger than that of TMDCs. Since the electronic properties of BP is thickness-dependent and changes with the numbers of layers, the optical band gaps of monolayer, bilayer and bulk BP are very different with values of 1.73 eV, 1.15 eV and 0.35 eV, respectively \cite{L. Li_2}. Unfortunately, monolayer and few-layer of BP are chemically unstable at ambient conditions, which is a critical obstacle for their practical applications.

Recently, a new family of 2D materials, group-IV metal monochalcogenides (also known as phosphorene analogues) with chemical formula MX (M = Ge , Sn ; X = S , Se ) have been extensively explored due to the low toxicity, earth abundant and chemical stability features \cite{L. C. Gomes,L. C. Gomes_2,A. S. Sarkar}. As one of group-IV metal monochalcogenides, bulk SnS adopts orthorhombic $\alpha$ phase structure at room temperature. The few-layer 2D SnS presents interesting properties like large absorbance coefficient, layer-dependent electronic structure, high in-plane anisotropy, odd-even quantum confinement effect, strong piezoelectricity and high carrier mobility \cite{A. S. Sarkar,C. Xin}. As a remarkable achievement, in 2015, Brent et al. synthesized bilayer SnS by liquid-phase exfoliation \cite{J. R. Brent}. Bilayer SnS which is a semiconductor with an indirect band gap \cite{L. C. Gomes} has been predicted to transform into a direct band gap semiconductor under a small in-plane uniaxial tensile strain \cite{Z.-Y. Li}.In addition, it has been found that by applying an electric field, the band gap of bilayer SnS decreases quickly and may vanish at some value.    

In 2013, the idea of hybrid van der Waals (vdW) heterostructure was introduced by Geim and Grigorieva \cite{A. K. Geim_2}, to highlight the advantages of stacking different monolayer or few-layer of 2D  materials via vdW interactions. This strategy is an excellent way to integrate and enhance electronic and optical properties of the individual components. Owing to the weakness of vdW interaction, the electronic properties of individual layers are maintained, and more importantly, new physical properties are emerged at the interface. Vertical vdW hybrid structures based on recently discovered 2D materials have a vast variety of applications in the design of next-generation electronic, spintronic and optoelectronic devices. Nowadays, several vdW heterostructures have been predicted theoretically or fabricated experimentally, such as graphene/phosphorene \cite{J. E. Padilha}, graphene/BN \cite{Y. Fan}, graphene/arsenene \cite{C. Xia}, graphene/SnS \cite{W. Xiong}, InSe / InTe \cite{J. Shang}, graphene/Mo$\textrm{S}_{2}$ \cite{M. Bernardi}, Mo$\textrm{S}_{2}$/Mo$\textrm{Se}_{2}$ \cite{F. Ceballos}, graphene/W$\textrm{Se}_{2}$ \cite{P. T. T. Le}, graphene/GeC \cite{T. V. Vu} and graphene/Zr$\textrm{S}_{2}$ \cite{S. Ur Rehman} . Recently, different types of vertical graphene-based vdW field effect transistors (FETs) have been demonstrated by using distinct electronic properties of 2D materials, like gate-tunable interface Schottky barrier which enables transition between tunneling and thermionic processes in such transistors \cite{S.-J. Liang,Z. Bai} . In fact, using graphene instead of metal as the contact can effectively alter the performance of devices based on heterostructures \cite{T. Roy}. For metal-semiconductor vdW hybrid structures, two types of contacts are formed: Schottky contact and Ohmic contact.  According to Schottky-Mott model, the n-type (p-type) Schottky barrier height (SBH), $\Phi_{Bn}$ ($\Phi_{Bp}$), is defined as $\Phi_{Bn} = E_{C} - E_{F}$ ($\Phi_{Bp} = E_{F} - E_{V}$) where $E_{C}$, $E_{V}$ and $E_{F}$ are the conduction band minimum, valence band maximum and Fermi level, respectively. In the case of Schottky contact (n-type or p-type), the Schottky barrier impedes efficient charge carrier injection and causes contact resistance. Fortunately, it is possible to tune SBH or contact type by an external electric field, a key factor which is crucial in designing of new electronic devices. Particularly, by applying a perpendicular electric field, a change from the p-type to n-type Schottky contact in both monolayer graphene/SnS and bilayer graphene/SnS heterostructures and a transition from Schottky contact to Ohmic contact in bilayer graphene/SnS heterostructure can be occurred \cite{W. Xiong}. 

Motivated by the successful exfoliation of bilayer SnS and fascinating properties of graphene/SnS vdW heterostructures, in this paper, we investigate the structural and electronic properties of the bilayer SnS/graphene, bilayer SnS/bilayer graphene (AA-stacked), bilayer SnS/bilayer graphene (AB-stacked) and monolayer SnS/graphene/monolayer SnS van der Waals (vdW) heterosructures by first-principles calculations. The intrinsic electronic properties of all components of these vdW heterostructures are well maintained. Moreover, we explore the effect of applying the external electric field on band structures of these hybrid 2D materials and then predict the variations of the interface contacts. Our results show that the electronic properties can be easily modulated by the perpendicular electric field, which is essential in nanoelectronic device applications. In addition, we find that the contact type at the interface will change by tuning the external electric field.

The rest of this paper is organized as follows: In Sec. II, we introduce the computational details of our calculations. In Sec. III, we present the electronic properties of the vdW heterostructures and the effects of the vertical electric field on their properties. Finally, in Sec. IV, we summarize our main findings.   

\section{computational details} 
We perform the first-principles calculations based on the density functional theory, as implemented in the QUANTUM ESPRESSO \cite{P. Giannozzi_1,P. Giannozzi_2} code to obtain the structural and electronic properties of the vdW heterostructures. The generalized gradient approximation (GGA) was used to describe exchange-correlation energy, as proposed by Perdew-Burke-Ernzerhof (PBE) \cite{J. P. Perdew} . Although PBE underestimates the band gap and SBH compared to HSE hybrid functional \cite{J. Heyd} , it is good at predicting the physical mechanisms and can give reasonable trends of the p-type and n-type SBH under the external electric field and strain\cite{C. Xia,W. Xiong,K. D. Pham} . The projector augmented wave (PAW) \cite{P. E. Blochl} potential was used to include the ion-electron interactions. In order to correctly describe vdW interaction, the DFT-D3 approach with Becke-Johnson damping proposed by Grimme \cite{S. Grimme_1,S. Grimme_2} was employed and a plane-wave cutoff energy of 50 Ry was used in the computations. The Brillouin zone was sampled with a $6\times14\times1$ Monkhorst-Pack grid \cite{H. J. Monkhorst} . All the geometric structures were optimized until the force on each atom was less than 0.01 eV/\AA . A vacuum space of 20 \AA \hspace{0.1 cm} along $z$ direction was included to avoid interaction between periodic images. Moreover, ab initio molecular dynamics (AIMD) simulations were performed by the Nose-Hoover method \cite{S. Nose_1,S. Nose_2,G. H. Hoover} using the OpenMX code \cite{T. Ozaki_1,T. Ozaki_2,T. Ozaki_3}.

\begin{figure*} 
	\includegraphics [width=1.0\linewidth,height=12cm]{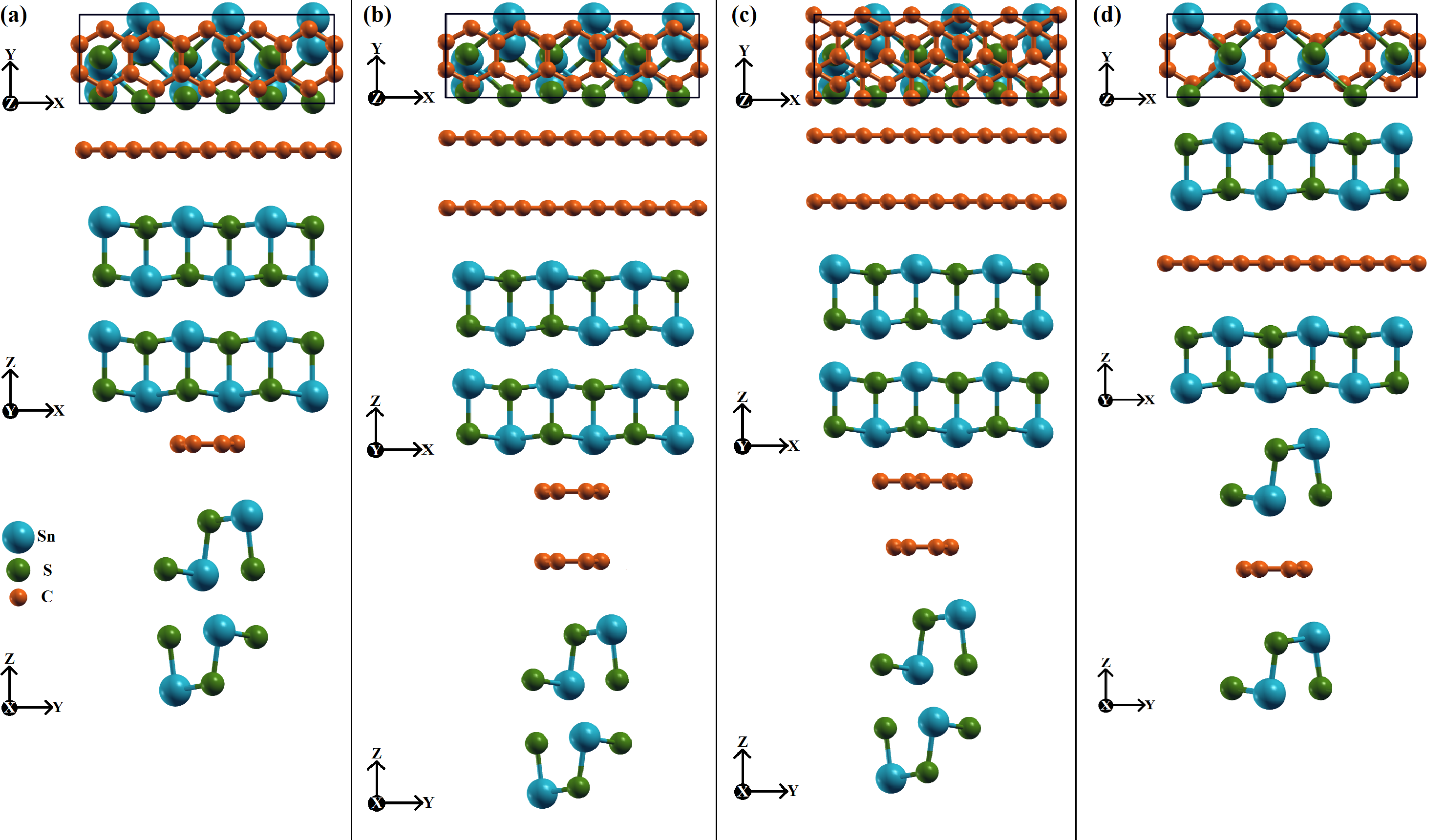} 
	\centering 
	\caption{Top and side views along $y$ and $x$ directions of the crystal structure of (a) bilayer SnS/graphene, (b) bilayer SnS/bilayer graphene (AA-stacked), (c) bilayer SnS/bilayer graphene (AB-stacked) and (d) monolayer SnS/graphene/monolayer SnS vdW heterostructures.}  
	
	\centering
	\label{fig:crystal_structures}
\end{figure*}  

\begin{table*}
	\renewcommand{\arraystretch}{1.2}
	\centering
	\setlength{\tabcolsep}{10pt}
	\caption{Binding energies and equilibrium interlayer distances }
	\label{tab-1}
	\begin{tabular}{ c c c}

		\hline
		\hline
		& Binding energy (meV) & Interlayer distance (\AA)  \\
		\hline
		Bilayer SnS/graphene & -39 & 3.49  \\
		
		Bilayer SnS/bilayer graphene (AA-stacked) & -39 & 3.48 \\
		
		Bilayer SnS/bilayer graphene (AB-stacked) & -41 & 3.49\\
		
		Monolayer SnS/graphene/monolayer SnS & -75 & 3.47 \\
		\hline
		\hline
	\end{tabular}
	
\end{table*}

\section{Results and discussion}
\subsection{Crystal structure, stability and electronic properties}
\begin{figure} [ht!]
	\includegraphics [width=1.0\linewidth]{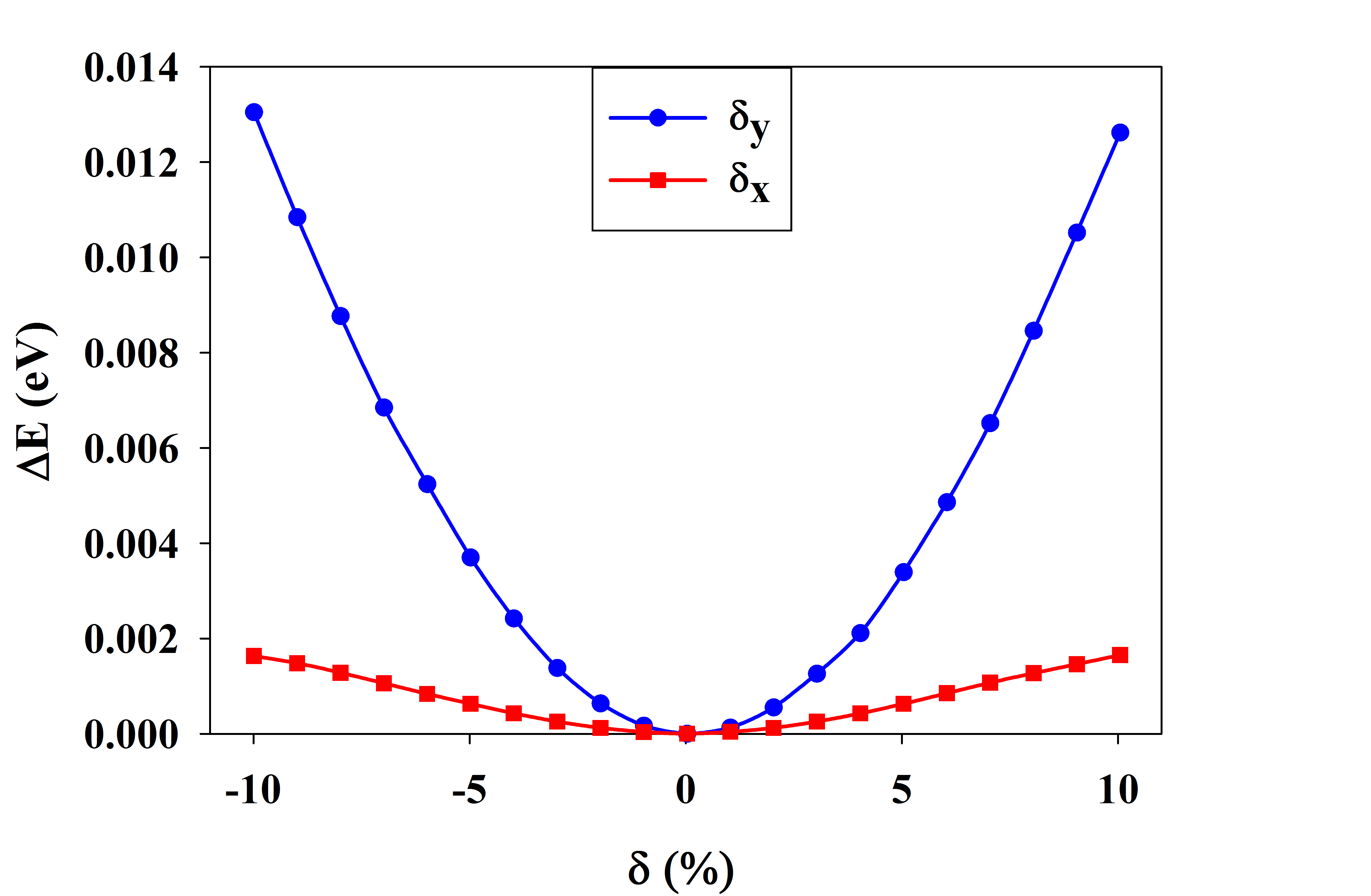} 
	\centering 
	\caption{Evolution of the total energy of bilayer SnS/graphene vdW heterostructure as a function of finite displacements of the bilayer SnS relative to graphene along $x$ ($\delta_{x}$) and $y$ ($\delta_{y}$) directions. }
	
	\centering
	\label{fig:Stacking_Pattern}
\end{figure}

In Figs. \ref{fig:crystal_structures}(a)-(d), we show top and side views of the optimized crystal structures of the bilayer SnS/graphene, bilayer SnS/bilayer graphene (AA-stacked), bilayer SnS/bilayer graphene (AB-stacked) and monolayer SnS/graphene/monolayer SnS vdW heterostructures, respectively.
The unit cell of the vdW heterostructure is consisted of $1\times5$ unit cells of  monolayer or bilayer graphene and $1\times3$ unit cells of monolayer or bilayer SnS. Due to the high sensitivity of 2D SnS properties to the strain, the bilayer SnS lattice is fixed and the monolayer, AA- or AB-stacked bilayer graphene lattice is compressed. The lattice mismatches in heterostructures with bilayer (monolayer) SnS along the $x$ and $y$ directions are less than 1.8\% and 1.2\% (1.1\% and 0.3\%), respectively which are small enough and within the acceptable limits for all the heterostructures to prevent any significant influence on the electronic properties. The calculated equilibrium interlayer distance for bilayer SnS is obtained 2.70 \AA \hspace{0.1 cm} in the vdW heterostructures studied here.
Also, the equilibrium interlayer distance between bilayer SnS and graphene is 3.49 \AA \hspace{0.1 cm}, similar to that of bilayer phosphorene/graphene heterostructure \cite{J. E. Padilha}. To find the configuration with minimum energy, we chose several starting positions of SnS layer relative to graphene (monolayer or bilayer) and relaxed the structures. Then, by moving 2D SnS relative to graphene along the $x$ ($\delta_{x}$) and $y$ ($\delta_{y}$) directions, we ensured that the minimum energy configurations were found.          	     		      	     		  
Fig. \ref{fig:Stacking_Pattern} depicts evolution of the total energy difference as a function of finite displacements $\delta_{x}$ and $\delta_{y}$ in bilayer SnS/graphene hybrid structure. As can be seen, the change in energy along the $y$ direction is greater than the $x$ direction (the same behavior was also observed for other vdW heterostructures studied here). However, these finite displacements do not affect the electronic properties of the vdW heterostructure, since the corresponding energy changes are quite small.

\begin{figure} [ht!]
	\includegraphics [width=1.0\linewidth,height=20cm]{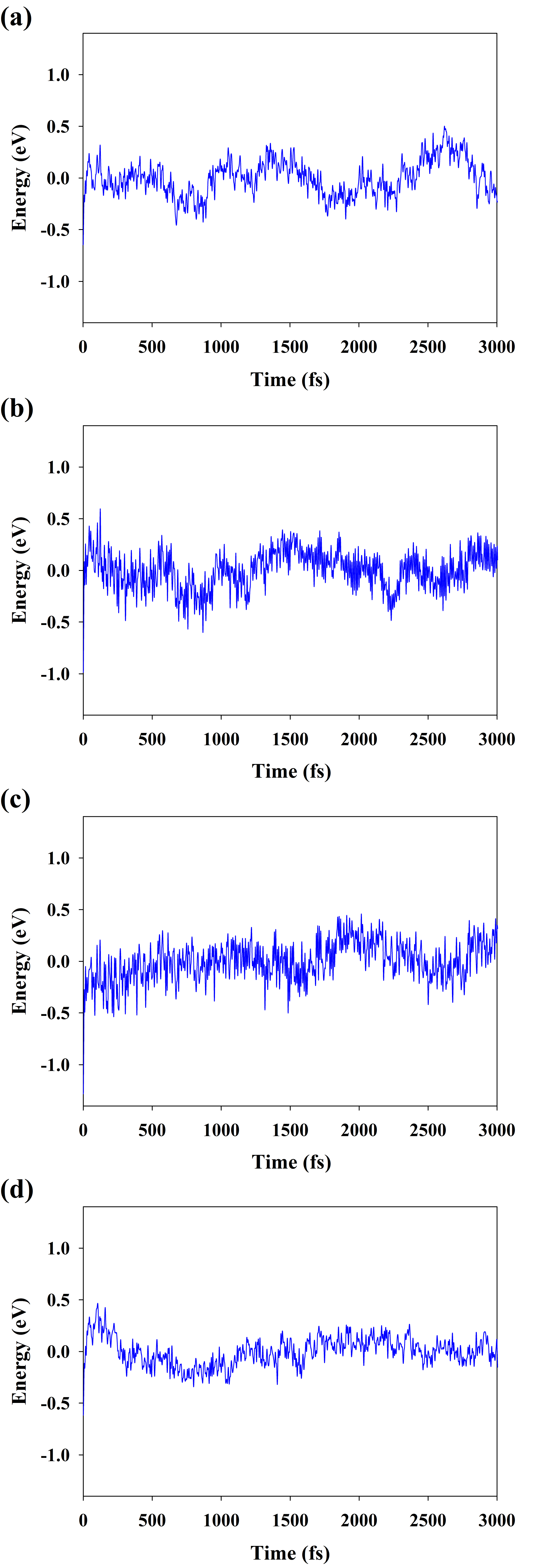}
	\centering  
	\caption{Fluctuations of total energies of (a) bilayer SnS/graphene, (b) bilayer SnS/bilayer graphene (AA-stacked), (c) bilayer SnS/bilayer graphene (AB-stacked) and (d) monolayer SnS/graphene/monolayer SnS vdW heterostructures as a function of time step.}
	
	\centering
	\label{fig:fluctuation_of_total_energy}
\end{figure}

In order to evaluate the stability of the vdW heterostructure, the binding energy of the system per carbon atom is calculated as {$E_{b} =[E_{vdW}-E_{SnS/SnS}-E_{MLG(BLG)}]/N$} in the cases of bilayer SnS/graphene and bilayer SnS/bilayer graphene (AA-stacked or AB-stacked) where $E_{vdW}$, $E_{SnS/SnS}$, $E_{MLG}$ and $E_{BLG}$ are the energies of vdW heterostructure, bilayer SnS, monolayer graphene and bilayer graphene, respectively and $N = 20$ is the number of carbon atoms in the vicinity of the bilayer SnS  in the unit cell. For monolayer SnS/graphene/monolayer SnS vdW heterostructure, the binding energy is obtained from {$E_{b} =[E_{vdW}-2E_{SnS}-E_{MLG}]/N$} with $E_{SnS}$ being the energy of monolayer SnS. The calculated binding energies and equilibrium interlayer distances between the components of heterostructures are listed in Table \ref{tab-1}. These values compared with those of other vdW heterostructures like  WS$_{2}$/graphene heterostructure (53 meV) \cite{T. P. Kaloni} and bilayer phosphorene/graphene heterostructure (60 meV) \cite{J. E. Padilha}, suggest that all proposed vdW heterostructures are energetically stable. Moreover, in order to check the thermal stability of the vdW heterostructures, we perform AIMD simulations at the temperature of 300 K for total 3 ps with a time step of 3 fs. Figs. \ref{fig:fluctuation_of_total_energy}(a)-(d) show the fluctuations  of total energies as a function of time step for bilayer SnS/graphene, bilayer SnS/bilayer graphene (AA-stacked), bilayer SnS/bilayer graphene (AB-stacked) and   monolayer SnS/graphene/monolayer SnS, respectively. As can be seen  the fluctuation  of total energies are small and there are no substantial structural deformation and bond breaking during the MD simulations.


\begin{figure*} [ht!]
	\includegraphics [width=1.0\linewidth,height=19cm]{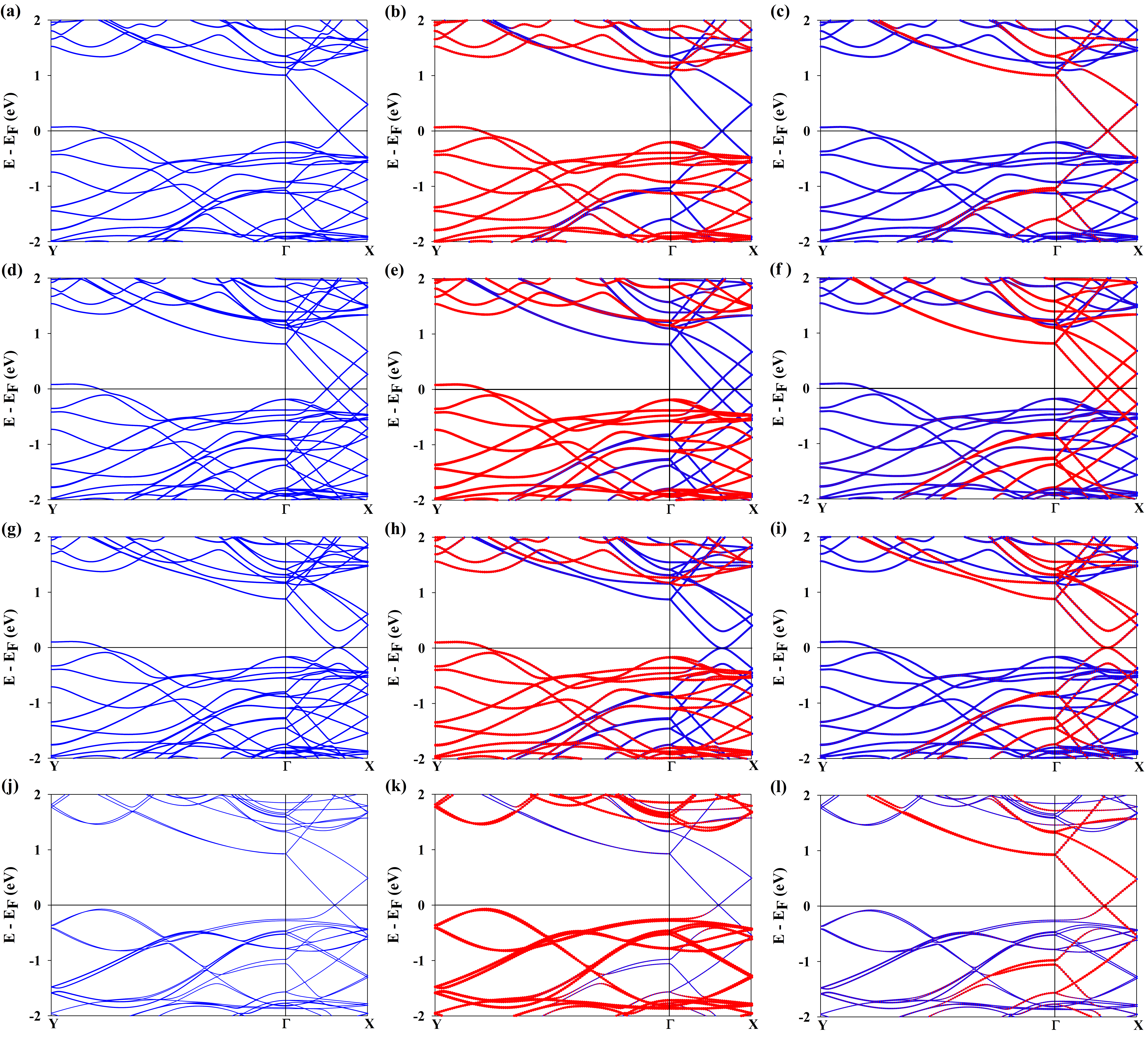} 
	\centering 
	\caption{ (a) Band structure of bilayer SnS/graphene vdW heterostructure and projected ones on (b) bilayer SnS and (c) graphene, (d) band structure of bilayer SnS/bilayer graphene (AA-stacked) vdW heterostructure and projected ones on (e) bilayer SnS and (f) bilayer graphene (AA-stacked), (g) band structure of bilayer SnS/bilayer graphene (AB-stacked) vdW heterostructure and projected ones on (h) bilayer SnS and (i) bilayer graphene (AB-stacked) and (j) band structure of monolayer SnS/graphene/monolayer SnS and projected ones on (k) both SnS monolayers and (l) graphene. (The projected bands are plotted by red dots.) }
	
	\centering
	\label{fig:bands_all}
\end{figure*} 

\begin{figure*} [ht!]
	\includegraphics [width=1.0\linewidth,height=20cm]{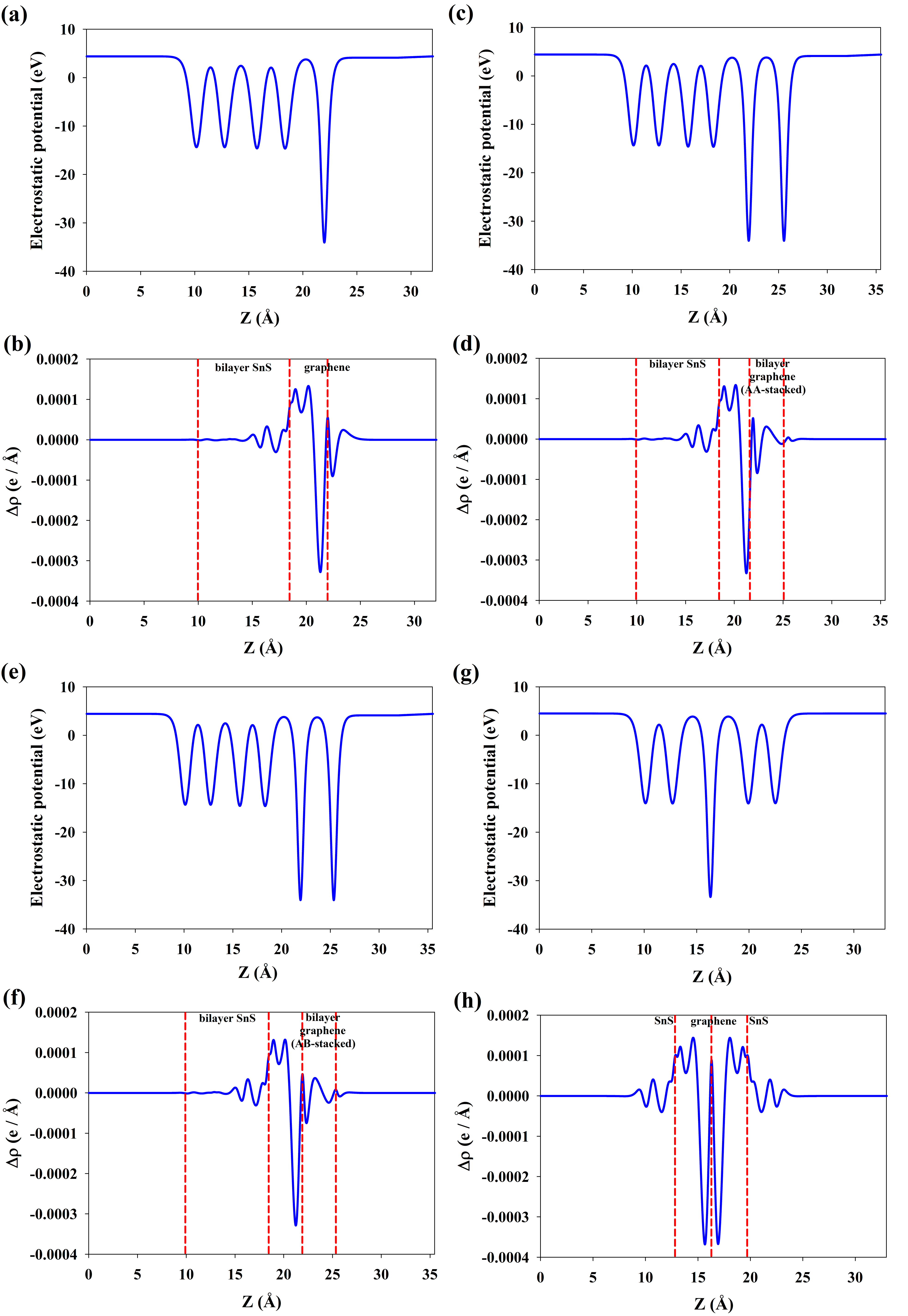} 
	\centering 
	\caption{ Electrostatic potential and plane-averaged charge density difference of (a) and (b) bilayer SnS/graphene, (c) and (d) bilayer SnS/bilayer graphene (AA-stacked), (e) and (f) bilayer SnS/bilayer graphene (AB-stacked) and (g) and (h) monolayer SnS/graphene/monolayer SnS vdW heterostructures along the $z$ direction. }
	
	\centering
	\label{fig:epotential_cdifference}
\end{figure*}           

In Fig. \ref{fig:bands_all}, the calculated band structures and projected band structures of all vdW heterostructures are illustrated. The band structure of the bilayer SnS/graphene vdW heterostructure and projected ones on the bilayer SnS and graphene are plotted in Figs. \ref{fig:bands_all}(a)-(c), respectively. As it can be observed, the band structure is a combination of its individual components in such a way that the band structure of bilayer SnS \cite{L. C. Gomes} and the Dirac cone of graphene are well maintained in the vdW heterostructure. 
In Figs. \ref{fig:bands_all}(d)-(f), the calculated band structure of the bilayer SnS/bilayer graphene (AA-stacked) vdW heterostructure and projected on the bilayer SnS and bilayer graphene (AA-stacked) are depicted, respectively. They clearly demonstrate that the band structure of the bilayer SnS and two Dirac cones of the bilayer graphene (AA-stacked) are well preserved in this heterostructure pointing out the weakness of vdW interaction. Also, according to Figs. \ref{fig:bands_all}(g)-(i), the electronic structure of the bilayer SnS and two pairs of parabolic bands of the AB-stacked bilayer graphene are retained in bilayer SnS/bilayer graphene (AB-stacked) hybrid structure, thus the intrinsic properties of the vdW heterostructure components are well preserved upon stacking. In the case of monolayer SnS/graphene/monolayer SnS vdW heterostructure, the calculated band structure and projected band structures on monolayer SnS and monolayer graphene shown in Figs. \ref{fig:bands_all}(j)-(l) again indicate that the original electronic properties of each component is maintained.
  
Based on the calculations presented above, all our hybrid structures exhibit two important aspects of vdW heterostructures; good stability and electronic properties preserving the main band structure characteristic of their individual constituents.

 \subsection{Interface contact type}
 
 It is well-known that the interface contact plays an important role in designing the electronic devices based on vdW heterostructures. So, we calculate the Schottky barrier heights for our 2D hybrid structures and determine their contact type. In bilayer SnS/graphene heterostructure the values of -0.07 eV and 1.10 eV are obtained for $\Phi_{Bp}$ and $\Phi_{Bn}$, respectively where the negative value of the $\Phi_{Bp}$ indicate that, unlike the monolayer SnS/graphene \cite{W. Xiong}, the bilayer SnS/graphene heterostructure forms an Ohmic contact. In addition, the calculated SBHs ($\Phi_{Bp}$ , $\Phi_{Bn}$) are (-0.09 eV , 1.12 eV) for bilayer SnS/AA-stacked bilayer graphene and (-0.11 eV , 1.13 eV) for bilayer SnS/AB-stacked bilayer graphene vdW heterostructures. Again, the smaller and negative values of $\Phi_{Bp}$ reveal forming an Ohmic contact at the interface of these hybrid structures. 
 However, we can induce a Schottky contact under the external conditions. In the case of monolayer SnS/graphene/monolayer SnS vdW heterostructure $\Phi_{Bp}$ and $\Phi_{Bn}$ have 0.07 eV and 1.34 eV values, respectively. Here, the value of $\Phi_{Bp}$ is smaller than $\Phi_{Bn}$, and thus, like monolayer SnS/graphene vdW heterostructure \cite{W. Xiong}, a p-type Schottky interface contact is formed which can be turn into an Ohmic one by applying a perpendicular electric field. 
 
 In order to investigate the charge transfer across the interface of bilayer SnS and monolayer (bilayer) graphene, we calculate the electrostatic potential and plane-averaged electron density difference, $\Delta\rho$, along the $z$ direction based on the following relation:  
\begin{equation}
	\begin{aligned}
		\Delta\rho(z) = \int\rho_{vdW}(x,y,z)dxdy -\int\rho_{SnS/SnS}(x,y,z)dxdy\\
		-\int\rho_{MLG(BLG)}(x,y,z)dxdy
	\end{aligned} 
\end{equation}
where $\rho_{vdW}$, $\rho_{SnS/SnS}$ and $\rho_{MLG (BLG)}$ are the charge density of the vdW heterostructure, bilayer SnS and monolayer (bilayer AA- or AB-stacked) graphene, respectively.        
For monolayer SnS/graphene/monolayer SnS heterostructure, we use $ \Delta\rho(z) = \int\rho_{vdW}(x,y,z)dxdy -\int\rho_{SnS(T)}(x,y,z)dxdy-\int\rho_{MLG}(x,y,z)dxdy-\int\rho_{SnS(B)}(x,y,z)dxdy $ with $\rho_{SnS(T)}(x,y,z) $ and $\rho_{SnS(B)}(x,y,z) $ being the charge density of the top and bottom  SnS monolayers, respectively. The results are shown in Figs. \ref{fig:epotential_cdifference}(a)-(h). It can be seen that the electrostatic potential is much deeper on the graphene (monolayer or bilayer) side, and as a result, this large electrostatic potential difference leads to an electric field across the interface which facilitates the charge transfer between components of the heterostructures. Also, we evaluate the work function, $W$, defined as the energy difference between vacuum level, $E_{vac}$, and Fermi energy, $ E_{F}$, for each hybrid structures and obtain the values of 4.41 eV, 4.43 eV, 4.45 eV and 4.50 eV for bilayer SnS/graphene, bilayer SnS/bilayer graphene (AA-stacked), bilayer SnS/bilayer graphene (AB-stacked) and monolayer SnS/graphene/monolayer SnS vdW heterostructures, respectively.  
Furthermore, according to calculations for the averaged charge density difference along $z$ direction displayed in Figs. \ref{fig:epotential_cdifference}(b),(d),(f) and (h), in all cases, charge is depleted  at graphene side and accumulated at SnS side; thus the charge transfers from monolayer or bilayer graphene to bilayer or monolayer SnS.

 \subsection{Electric field effect}
 
\begin{figure} [ht!]
	\includegraphics [width=1.0\linewidth]{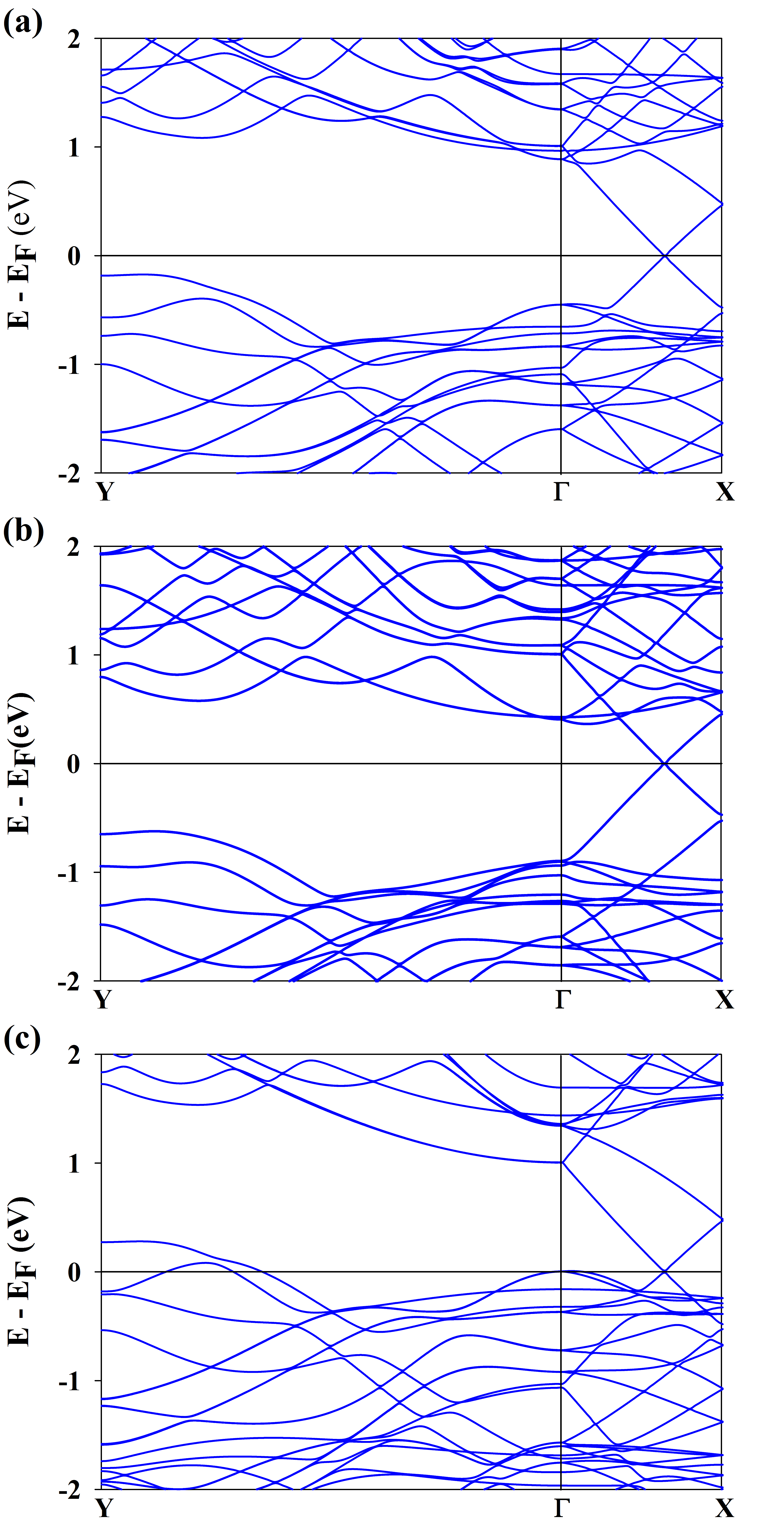} 
	\centering 
	\caption{ Band structure of bilayer SnS/graphene vdW heterostructure under the electric fields of (a) 0.1 V/\AA, (b) 0.3 V/\AA \hspace{0.1 cm} and (c) -0.1 V/\AA .
	}
	
	\centering
	\label{fig:bands_sns_sns_g_electric}
\end{figure} 
As previously stated, it is of crucial importance to control SBH for the electronic device applications. In particular, applying an external perpendicular electric field causes the evolution of band structure and, consequently, provides an efficient way to tune the SBH of graphene-based vdW heterostructures \cite{J. E. Padilha,C. Xu,Huynh V. Phuc,M. Sun}. It was shown that, in the case of graphene/phosphorene monolayer (bilayer), the zero field p-type Schottky contact changes to the n-type under an external field greater than 0.18 (0.2) V/\AA \hspace{0.1 cm} and for the negative values smaller than -0.18 V/\AA, turns to Ohmic one in graphene/phosphorene bilayer \cite{J. E. Padilha} . Also, when a negative electric field smaller than -0.125 V/\AA \hspace{0.1 cm} is applied to graphene/GeSe heterostructure, the p-type contact transforms to the Ohmic one \cite{C. Xu} . Furthermore, in graphene/GaSe bilayer, while for positive electric field larger than 0.025 V/\AA, a transition from n-type to p-type Schottky contact will occur, the contact type will not change under the negative field \cite{Huynh V. Phuc} . Figs. \ref{fig:bands_sns_sns_g_electric}(a)-(c) present the band structure of of the bilayer SnS/graphene under the electric fields 0.1 V/\AA, 0.3 V/\AA \hspace{0.1 cm} and -0.1 V/\AA, respectively. It can be realized that while the contact type is transformed from Ohmic to p-type Schottky one at the electric field of 0.1 V/\AA, the application of a negative electric field of -0.1 V/\AA \hspace{0.1 cm} does not change the contact type but shifts the valence band maximum (VBM) of bilayer SnS upward and as a result, the SBH decreases. According to Fig. \ref{fig:bands_sns_sns_g_electric}(b) under an electric field of 0.3 V/\AA, a transition from the p-type to n-type Schottky contact occurs in bilayer SnS/graphene vdW heterostructure.

\begin{figure} [ht!]
	\includegraphics [width=1.0\linewidth]{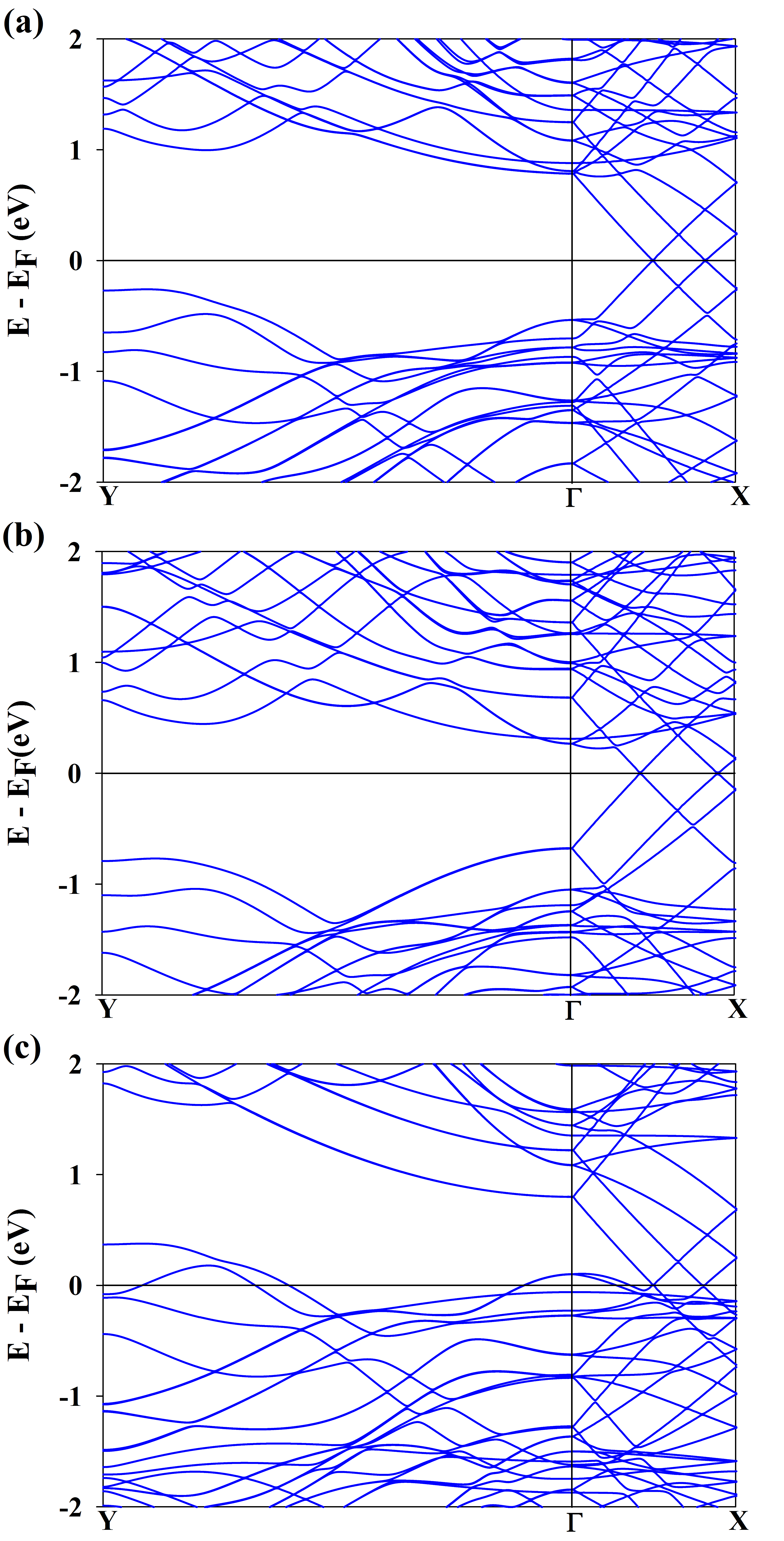} 
	\centering 
	\caption{ Band structure of bilayer SnS/bilayer graphene (AA-stacked) vdW heterostructure under the electric fields (a) 0.1 V/\AA, (b) 0.26 V/\AA \hspace{0.1 cm} and (c) -0.1 V/\AA.	}
	
	\centering
	\label{fig:bands_sns_sns_g_g(AA)_electric}
\end{figure}

\begin{figure} [ht!]
	\includegraphics [width=1.0\linewidth]{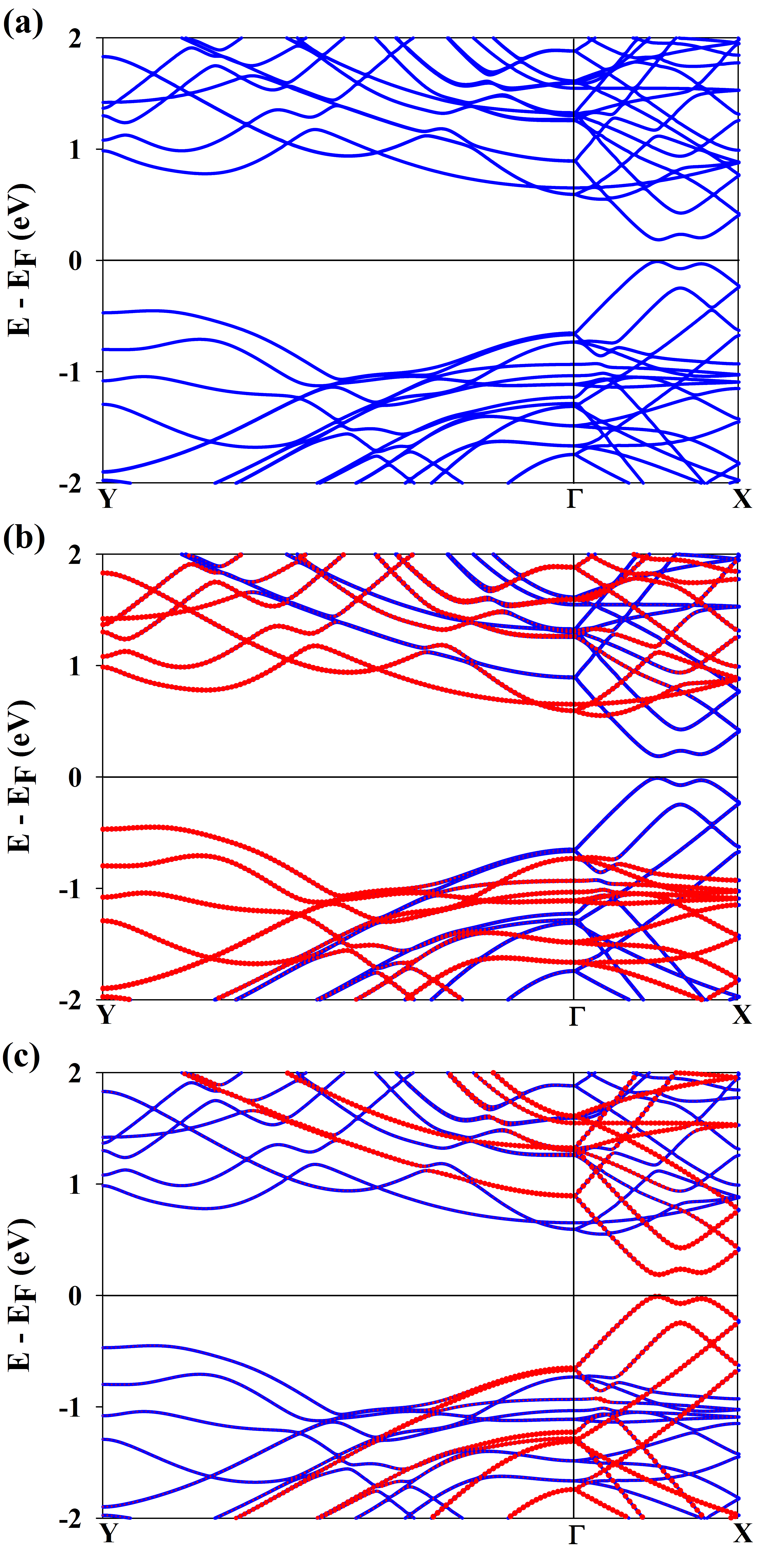} 
	\centering 
	\caption{(a) Band structure of bilayer SnS/bilayer graphene (AB-stacked) vdW heterostructure under the electric field 0.2 V/\AA.  Band structure of the vdW heterostructure projected on (b) bilayer SnS and (c) bilayer graphene (AB-stacked). }
	
	\centering
	\label{fig:band_sns_sns_g_g(AB)_electric}
\end{figure}

\begin{figure} [ht!]
 	\includegraphics [width=1.0\linewidth]{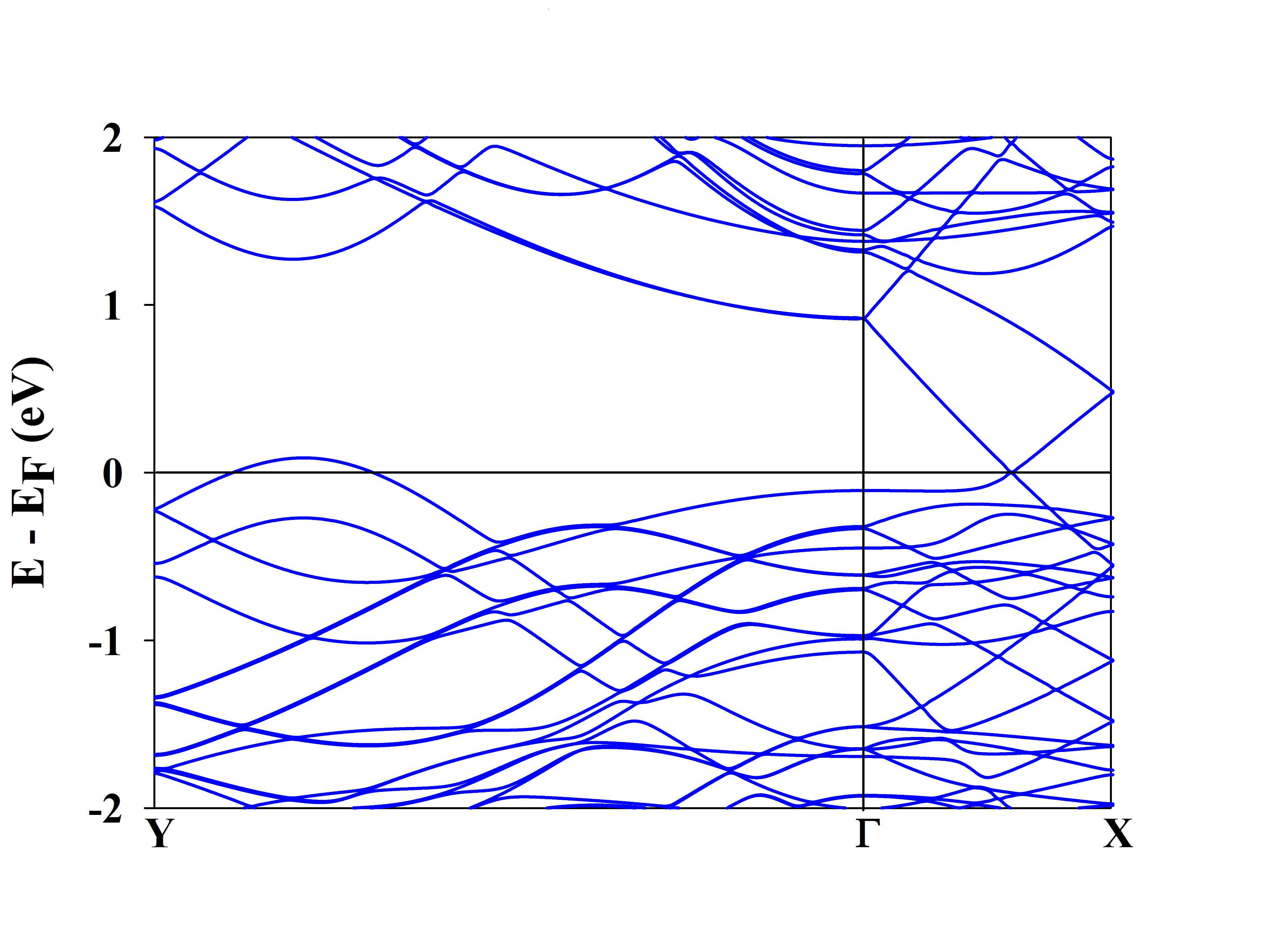} 
 	\centering 
 	\caption{Band structure of monolayer SnS/graphene/monolayer SnS vdW heterostructure under the electric field 0.1 V/\AA .}  
 	
 	\centering
 	\label{fig:band_sns_g_sns_electric}
 \end{figure}

Our calculations predict a very similar result for bilayer SnS/bilayer graphene (AA-stacked) in the presence of an external electric field (See Fig. \ref{fig:bands_sns_sns_g_g(AA)_electric}). In a field strength of 0.1 V/\AA \hspace{0.1 cm}, the contact type is transformed from Ohmic to p-type Schottky contact, at higher value of 0.26 V/\AA \hspace{0.1 cm}, a n-type contact is obtained and finally under a negative field (-0.1 V/\AA), still there is an Ohmic contact with upward shifted the VBM of bilayer SnS. 
Figs. \ref{fig:band_sns_sns_g_g(AB)_electric}(a)-(c) show the calculated band structure of the bilayer SnS/bilayer graphene (AB-stacked) vdW heterostructure and the band structures projected on the bilayer SnS and bilayer graphene (AB-stacked) under the electric field 0.2 V/\AA, respectively. Upon application of an external perpendicular electric field, the inversion symmetry of the bilayer graphene (AB-stacked) is broken \cite{Y. Zhang} and a Mexican hat band structure with a gap is formed in the low-energy spectrum, in agreement with previous studies \cite{H. Min,E. McCann,E. V. Castro}. Interestingly, it is possible to adjust this band gap by the external electric field; e.g. under an electric field of 0.2 V/\AA, the induced band gap will be about 0.21 eV. Therefore, applying the electric field leads to the transition from semiconductor/semimetal to semiconductor/semiconductor contact in bilayer SnS/bilayer graphene (AB-stacked) vdW heterostructure which enhances its optical absorption and excitonic effects and makes it a promising candidate for next-generation tunable electro-optical devices. 
For monolayer SnS/graphene/monolayer SnS hybrid structure, we calculate the band structure under an electric field of 0.1 V/\AA (Fig. \ref{fig:band_sns_g_sns_electric}) and find that the highest valence band and lowest conduction band of the two SnS monolayers are separated in such a way that the highest valence band crosses the Fermi level, leading to an Ohmic contact.
 \begin{figure} [ht!]
	\includegraphics [width=1.0\linewidth,height=16cm]{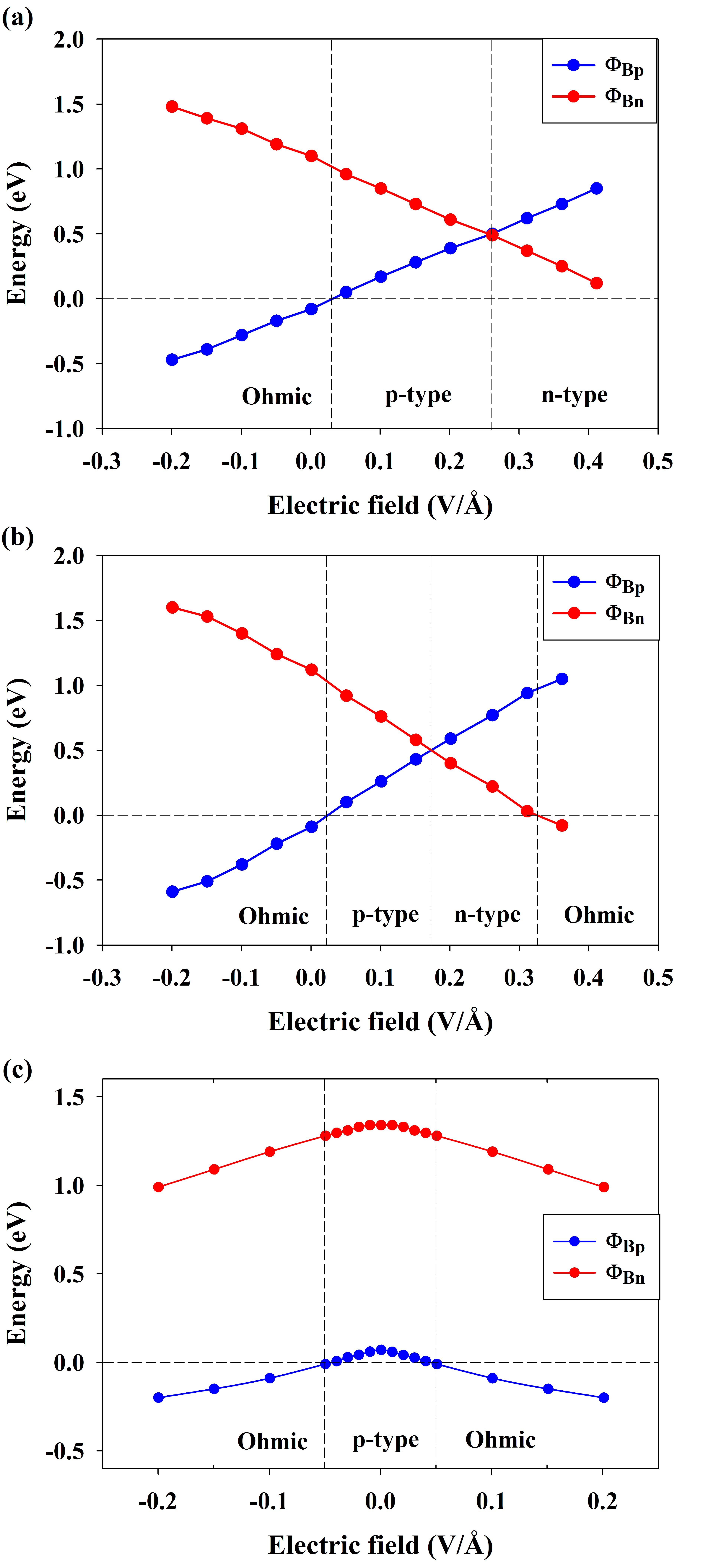}
	\centering  
	\caption{(a)-(c) Evolution of p-type and n-type SBH of bilayer SnS/graphene, bilayer SnS/bilayer graphene (AA-stacked) and monolayer SnS/graphene/monolayer SnS vdW heterostructures, respectively as a function of the external electric field.}
	
	\centering
	\label{fig:electric_field_evolution}
\end{figure}
 \begin{figure} [ht!]
	\includegraphics [width=1.0\linewidth,height=16cm]{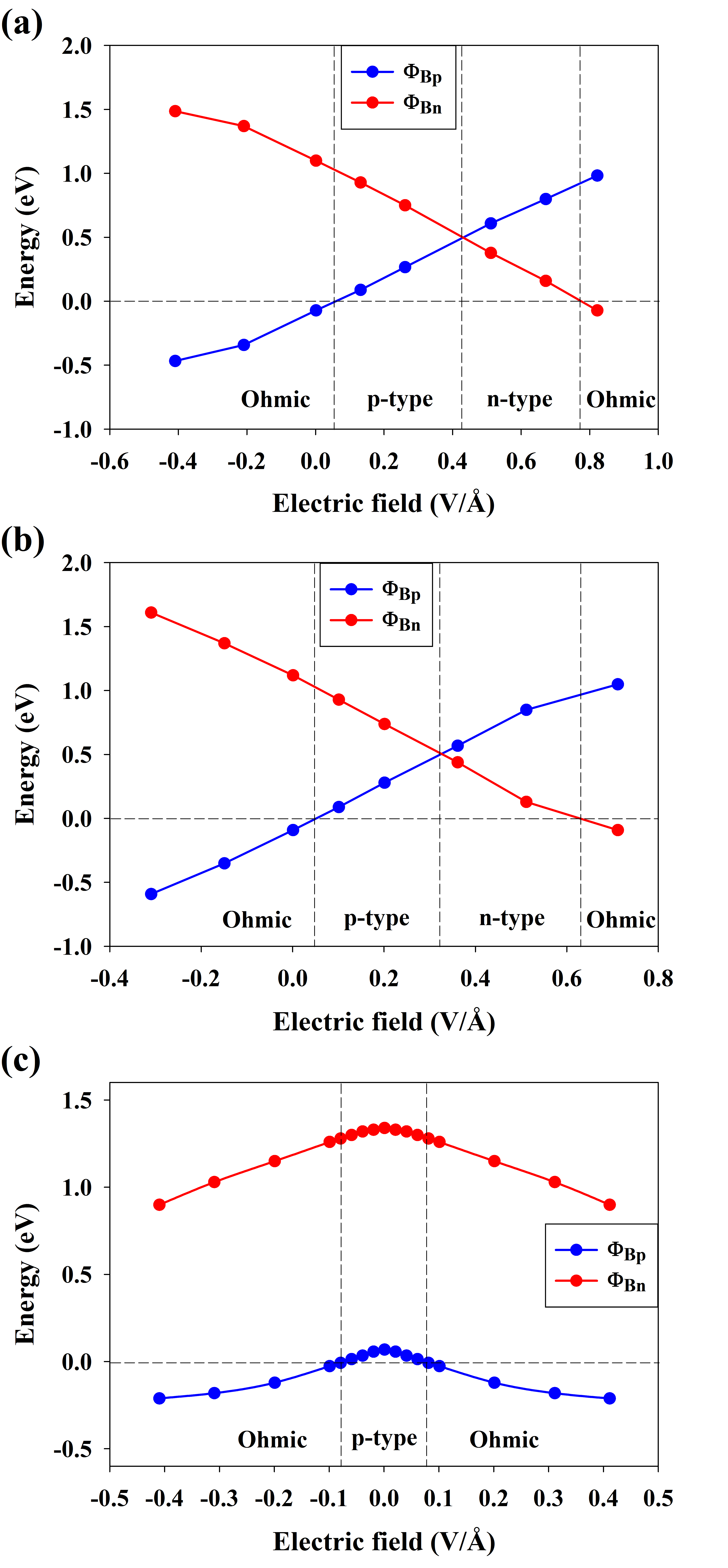}
	\centering  
	\caption{(a)-(c) Evolution of dipole-corrected p-type and n-type SBH of bilayer  SnS/graphene, bilayer SnS/bilayer graphene (AA-stacked) and monolayer SnS/graphene/monolayer SnS  vdW heterostructures, respectively as a function of the external electric field.}
	
	\centering
	\label{fig:electric_field_evolution2}
\end{figure}

In order to explore the interfacial contact property more accurately, we plot the evolution of contact type and the change of SBH as a function of the external electric field in Figs. \ref{fig:electric_field_evolution}(a)-(c). As can be seen from Figs. \ref{fig:electric_field_evolution}(a) and (b), the zero field Ohmic contact in both bilayer SnS/graphene and bilayer SnS/bilayer graphene (AA-stacked) heterostructures turns into the p-type Schottky one at fields of 0.03 V/\AA \hspace{0.1 cm} and 0.02 V/\AA, respectively. In addition, further increasing of the electric field leads to the decrease of n-type and increase of p-type SBHs and for even higher values, a transition from p-type to n-type Schottky contact occurs. On the other hand, for the negative values of electric field, the Ohmic contact persists but the p-type SBH reduces with decreasing the electric field strength. 
One can notice that in monolayer SnS/graphene/monolayer SnS vdW heterostructure, as illustrated in Fig. \ref{fig:electric_field_evolution}(c), when the electric field is in the range of -0.05 V/\AA \hspace{0.1 cm} and 0.05 V/\AA \hspace{0.1 cm}, the contact type remains p-type Schottky contact but for the electric fields greater (less) than 0.05 V/\AA (-0.05 V/\AA), the contact turns to Ohmic one. Moreover, unlike the other vdW heterostructures studied here, the positive and negative electric fields have the same effects on the n-type and p-type SBH due to the symmetric crystal structure of monolayer SnS/graphene/monolayer SnS vdW heterostructure along the $z$ direction.

In order to investigate the effect of the dipole correction on the SBH, we calculate the electronic structure of the vdW heterostructures with inclusion of  the dipole correction. Figs. \ref{fig:electric_field_evolution2}(a)-(c) show the evolution of p-type and n-type SBHs as a function of external electric field for  bilayer SnS/graphene, bilayer SnS/bilayer graphene (AA-stacked) and monolayer SnS/graphene/monolayer SnS vdW heterostructures, respectively when the dipole correction is included in the calculations. It can be observed that the general trends and features of p-type and n-type SBHs display similar behavior with and without the dipole correction, however, inclusion of  the dipole correction reduces the variation of SBH by external electric field, thereby the zero field Ohmic contact in the bilayer SnS/graphene and bilayer SnS/bilayer graphene (AA-stacked) heterostructures turns into the p-type Schottky one at larger fields of 0.05 V/\AA  \hspace{0.1 cm} and 0.04 V/\AA , respectively and in the case of the monolayer SnS/graphene/monolayer SnS  vdW heterostructure, the contact type remains p-type Schottky contact in the wider range of  -0.08 V/\AA  \hspace{0.1 cm} and 0.08 V/\AA .

\section{Conclusion}

In summary, using density functional theory, we have investigated the structural and electronic properties of the bilayer SnS/graphene, bilayer SnS/bilayer graphene (AA-stacked), bilayer SnS/bilayer graphene (AB-stacked) and monolayer SnS/graphene/monolayer SnS vdW heterostructures. We have shown that the intrinsic electronic properties of all components of the vdW heterostructures are well preserved upon stacking and the application of a perpendicular electric field is an efficient way to adjust the band structure and to control the contact properties.  In both bilayer SnS/graphene and bilayer SnS/bilayer graphene (AA-stacked) heterostructures, an interfacial Ohmic contact is formed which can easily transform to the p-type and then n-type Schottky contacts under the external electric field. On the other hand, although the vdW heterostructure of bilayer SnS/bilayer graphene (AB-stacked) is also an Ohmic interface, applying an external electric field leads to symmetry breaking in AB-stacked bilayer graphene and forming a Mexican hat dispersion in the low-energy spectrum of its band structure with a tunable band gap. Thus, a transition from semimetal/semiconductor contact to semiconductor/semiconductor contact occurs in this vdW heterostructure which can enhance its optical properties. Furthermore, calculations show that a p-type Schottky contact is created at the interface of monolayer SnS/graphene/monolayer SnS heterostructure and predict a transition from Schottky to Ohmic contact in the presence of both positive and negative external electric fields. Our results interestingly indicate that all of these vdW heterostructures have promising potential applications in nanoelectronic devices.

           	\end {document}